\documentclass[aps,prb,reprint,superscriptaddress,hyphens]{revtex4-1}

\usepackage{graphicx}
\usepackage{amsmath}

\begin{document}

\title{Magnetic anisotropy in antiferromagnetic hexagonal MnTe}

\author{D.~Kriegner}
\email[]{dominik.kriegner@gmail.com}
\affiliation{Institute of Physics, Academy of Science of the Czech Republic, Cukrovarnick\'a 10, 162 00 Praha 6, Czech Republic}
\affiliation{Charles University in Prague, Ke Karlovu 3, 121 16 Praha 2, Czech Republic}

\author{H.~Reichlova}
\affiliation{Institute of Physics, Academy of Science of the Czech Republic, Cukrovarnick\'a 10, 162 00 Praha 6, Czech Republic}

\author{J.~Grenzer}
\affiliation{Helmholtz-Zentrum Dresden-Rossendorf, Institute of Ion Beam Physics and Materials Research, Bautzner Landstrasse 400, 01328 Dresden, Germany}

\author{W. Schmidt}
\affiliation{J\"ulich Centre for Neutron Science JCNS, Forschungszentrum J\"ulich GmbH, Outstation at ILL, CS 20156, 71 avenue des Martyrs, F-38042 Grenoble, France}

\author{E. Ressouche}
\affiliation{Univ. Grenoble Alpes, CEA, INAC-MEM, 38000 Grenoble, France}

\author{J.~Godinho}
\affiliation{Institute of Physics, Academy of Science of the Czech Republic, 
Cukrovarnick\'a 10, 162 00 Praha 6, Czech Republic}

\author{T. Wagner}
\affiliation{Hitachi Cambridge Laboratory, Cambridge CB3 0HE, United Kingdom}

\author{S. Y. Martin}
\affiliation{Hitachi Cambridge Laboratory, Cambridge CB3 0HE, United Kingdom}
\affiliation{Univ. Grenoble Alpes, CNRS, CEA, Grenoble INP, INAC-Spintec, 
38000 Grenoble, France}

\author{A.B.~Shick}
\affiliation{Institute of Physics, Academy of Science of the Czech
  Republic, Na Slovance 1999/2, 182 21 Praha 8, Czech Republic}

\author{V.~V.~Volobuev}
\affiliation{Institute of Semiconductor and Solid State Physics, Johannes Kepler University Linz, Altenbergerstr. 69, 4040 Linz, Austria}
\affiliation{National Technical University, ''Kharkiv Polytechnic Institute'', 61002 Kharkiv, Ukraine}

\author{G.~Springholz}
\affiliation{Institute of Semiconductor and Solid State Physics, Johannes Kepler University Linz, Altenbergerstr. 69, 4040 Linz, Austria}

\author{V.~Hol\'y}
\affiliation{Charles University in Prague, Ke Karlovu 3, 121 16 Praha 2, Czech Republic}

\author{J. Wunderlich}
\affiliation{Hitachi Cambridge Laboratory, Cambridge CB3 0HE, United Kingdom}

\author{T.~Jungwirth}
\affiliation{Institute of Physics, Academy of Science of the Czech Republic, Cukrovarnick\'a 10, 162 00 Praha 6, Czech Republic}
\affiliation{School of Physics and Astronomy, University of Nottingham, Nottingham NG7 2RD, United Kingdom}

\author{K.~V\'yborn\'y}
\affiliation{Institute of Physics, Academy of Science of the Czech Republic, Cukrovarnick\'a 10, 162 00 Praha 6, Czech Republic}

\date{\today}

\begin{abstract}
  Antiferromagnetic hexagonal MnTe is a promising material for spintronic
  devices relying on the control of antiferromagnetic domain orientations. Here
  we report on neutron diffraction, magnetotransport, and magnetometry
  experiments on semiconducting epitaxial MnTe thin films together with density
  functional theory (DFT) calculations of the magnetic anisotropies.
  The easy axes
  of the magnetic moments within the hexagonal basal plane are determined to be
  along $\left<1\bar100\right>$ directions. The spin-flop transition and
  concomitant repopulation of domains in strong magnetic fields is observed.
  Using epitaxially induced strain the onset of the spin-flop transition
  changes from $\sim2$~T to $\sim0.5$~T for films grown on InP and SrF$_2$
  substrates, respectively.
\end{abstract}

\maketitle

\section{Introduction}

Antiferromagnets (AFMs) have recently attracted considerable attention in
the context of spintronic devices\cite{Jungwirth2016, Baltz2017} not
only as passive
components (e.g. pinning layers in magnetic tunnel junctions) but also directly
as a medium to store information.\cite{Olejnik2017} One of the key requirements
for magnetically ordered materials to provide device functionality is the
possibility to manipulate the magnetic moments. Albeit not straightforward for
AFMs (at least in comparison to ferromagnets), this turns out to be possible in
several ways~\cite{Marti2014, Reichlova2015, Wadley2016, Kriegner2016}. For
example spin orbit torques either induced by interfaces\cite{Reichlova2015} or
by the inverse spin galvanic effect inside the antiferromagnet\cite{Wadley2016}
can be used to manipulate antiferromagnetic states. Further, field cooling
through the N\'eel temperature\cite{Marti2014, Kriegner2016} or high magnetic
fields applied in the antiferromagnetic state\cite{Kriegner2016} were shown to
manipulate the domain population. For this purpose, as well as for detection
and stability of ordered magnetic states, magnetic anisotropies (MAs)
have to be considered carefully, which is the aim of this work.

One of the main advantages of using AFMs instead of ferromagnets in
spintronics is the availability
of a broad variety of intrinsic antiferromagnetic semiconductors.  They allow
for merging the vast amount of spintronic effects with electrical controlled
transport properties of a semiconductor.  Among them, hexagonal manganese
telluride (MnTe) had already been extensively studied well before the advent of
spintronics, in nineteen-sixties and -seventies, for its
optical\cite{Onari1974} or magnetical\cite{Komatsubara1963} properties.  It has
a relatively high N\'eel temperature ($T_{\rm N}$) of 310~K\cite{LB_MnTe} and a
moderate band gap of 1.27 to 1.46~eV.\cite{Ferrer-Roca2000,Kriegner2016}
Typically MnTe is a $p$-type semiconductor but intentional doping with sodium
or chromium can tune the resistivity over a wide range.\cite{WasscherPhd} The
magnetic structure of MnTe was determined from neutron diffraction and consists
of ferromagnetic hexagonal Mn planes which are antiferromagnetically coupled
along the
$c$-axis.\cite{Kunitomi1964, Efrem2005,Szuszkiewicz2005,Przezdziecka2008}
The determination of magnetic moment of Mn atoms was a subject of
several experimental works (see the summary in Ref.~\onlinecite{Kunitomi1964}),
the scatter being relatively large (from values close to 5$\mu_B$
all the way down to almost 4$\mu_B$) and moreover,
the easy axis was not determined.  Neither spin-wave measurements by
inelastic neutron diffraction could resolve the MA within the hexagonal
$c$-plane~\cite{Szuszkiewicz2006} although this MA is certainly
present as implied by
torque magnetometry\cite{Komatsubara1963} and recent anisotropic
magneto-resistance measurements~\cite{Kriegner2016}. Both measurements have
shown that three distinct orientations of domains exist within the hexagonal
plane. The possibility to change the domain population almost
continuously\cite{Kriegner2016} also affords memristive
behavior\cite{Chua1971,Meuffels2012,Lequeux2016} to MnTe-based devices.

Although bulk materials were explored first, the attention turned later to thin
layers grown on various substrates, leading to the discovery of a
new MnTe phase.
Apart from the hexagonal phase (NiAs-type, $\alpha$-phase, P6$_3$/mmc (194),
Fig.~\ref{fig:1}(a)) which is stable in bulk form, zinc-blende MnTe films
with a larger optical band gap and much lower N\'eel temperature\cite{Ando1992}
was found to be epitaxially stabilized on GaAs substrates with and without CdTe
buffer layer\cite{Akinaga1993,Janik1995,Hennion2002}. Epitaxial thin film
growth of the hexagonal $\alpha$-MnTe phase, which is the topic of this work,
was demonstrated on single crystalline SrF$_2$(111),
InP(111),\cite{Kriegner2016} and Al$_2$O$_3$(0001)\cite{Przezdziecka2008,
Kim2009, Wang2014} substrates as well as on amorphous
Si(111)/SiO$_2$\cite{Wang2012}. Due to lattice and thermal expansion
coefficient mismatch between $\alpha$-MnTe and the substrates, films will
experience strain that may affect the magnetic properties such as MAs.
For example,\cite{Zemen2009} the dilute magnetic
semiconductor (Ga,Mn)As is known to have an in-plane MA under compressive
strain and an out-of-plane MA for tensile strain under suitable conditions.
Here we study the MAs in MnTe on different substrates, which cause different
strain states.  The knowledge of the easy axis directions is crucial for
transport phenomena modeling, which has so far relied only on
assumptions.\cite{Kriegner2016}  As far as the easy axis directions are
concerned, we confirm these assumptions using DFT+U calculations combined with
experiments. Using magneto-transport, magnetometry and neutron diffraction, we
determine the easy axes to be along $\left<1\bar100\right>$ and show in what
respect MAs are sensitive to epitaxy--induced strain.

The paper is organized as follows. After introduction of the results of DFT+U
calculations in Sec.~II, we describe our samples structure and basic
magnetometry characterization in Sec.~III. Section~IV presents our neutron
diffraction experiments; section~V complementary magneto-transport studies.
Further magnetometry experiments determining the spin-flop field are presented
in Sec.~VI. Finally, we conclude in Sec.~VII.

\section{Magnetic anisotropy calculations}

The magnetic anisotropy energy (MAE) in antiferromagnets comprises two main
contributions: the dipole term and the magnetocrystalline anisotropy (MCA). In
order to calculate the latter, we use the relativistic version of the
rotationally invariant DFT+U method\cite{Shick2001} which takes into account
spin-orbit coupling, and non-diagonal in spin contributions into the occupation
matrix. The full-potential linearized augmented plane-wave
(FLAPW)\cite{Wimmer1981} basis is used in the self-consistent total energy
calculations. We use $U=4$~eV and $J=0.97$~eV parameters taken from a similar
compound of manganese~\cite{Antropov2014}.

\begin{figure}
\includegraphics[width=\linewidth]{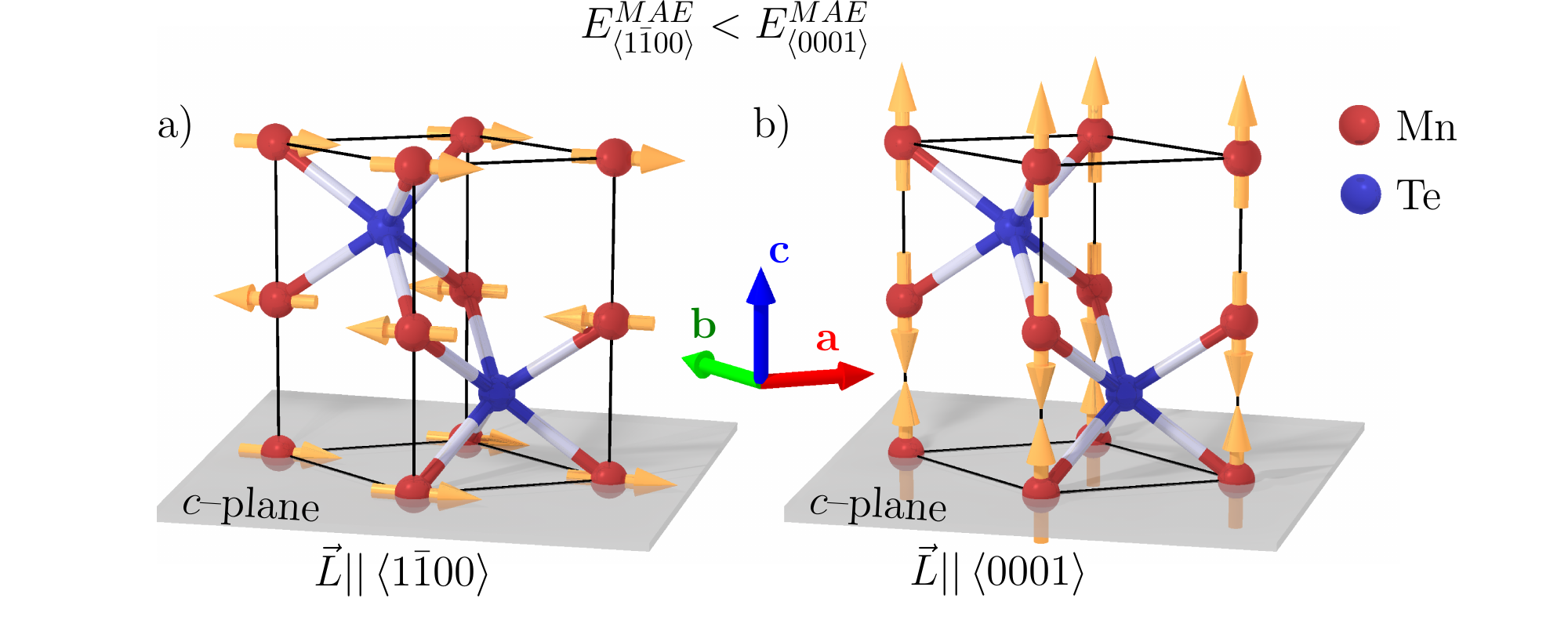}
\caption{Sketch of the atomic and possible magnetic structures of
antiferromagnetic hexagonal MnTe. (a) In-plane/$c$-plane (ground state) and (b)
out-of-plane/$c$-axis (hard axis) orientation of the magnetic moments of Mn
with the N\'eel vector $\vec L$ along $\left<1\bar100\right>$ and
$\left<0001\right>$ are shown. The hexagonal basal plane, i. e. the $c$-plane
is indicated by a gray plane, while red, green, and blue arrows show the
directions of the unit cell axes.}
\label{fig:sketch}\label{fig:1}
\end{figure}

The dipole term is a classical contribution from dipole-dipole interaction of
localized magnetic moments \cite{Wien2k}. For coherent rotations of the two AFM
sublattices which strictly maintain their antiparallel alignment, e.g. one
which interpolates between the two magnetic configurations shown in
Fig.~\ref{fig:sketch}, the dipole term depends in general on the rotation
angle. This dependence is absent for cubic crystals but present in MnTe since
the crystal symmetry of the NiAs-structure is lower. This causes that the
energy of the dipole-dipole interaction of the structure in
Fig.~\ref{fig:sketch}(b), with magnetic moments aligned along the $c$-axis is
higher than that of any structure with magnetic moments oriented in the
hexagonal basal plane ($c$-plane), e.g.  Fig.~\ref{fig:sketch}(a). 

For lattice constants $a=0.4134$~nm and $c=0.6652$~nm (experimentally
determined at 5~K / see Sec.  III, Fig.~\ref{fig:xrdtemp}(a,b)), we obtain that
Mn atoms carry the magnetic moments of 4.27$~\mu_B$ (spin $M_S=4.25~\mu_B$ plus
orbital $M_L=0.02~\mu_B$ magnetic moments). The energy difference of the two
different configurations shown in Fig.~\ref{fig:sketch} from the dipole term
$E^{\mathrm{dipole}}$ is calculated to be 0.135~meV per unit cell, favoring the
alignment in the $c$-plane. This contribution to MAE is only weakly
dependent on strain or relevant lattice distortions and gives no anisotropy
within the $c$-plane.

The DFT+U calculations of the MCA are much more involved but, rather generally,
a clear picture emerges of moderately large out-of-plane anisotropy and small
anisotropies within the $c$-plane. For the lattice constants quoted above, an
energy difference between configurations in Fig.~\ref{fig:sketch}(a) and
Fig.~\ref{fig:sketch}(b) of $0.11$~meV per unit cell is calculated again
favoring the alignment in the $c$-plane. The anisotropy within the $c$-plane,
defined as the energy difference between the magnetic structure in
Fig.~\ref{fig:sketch}(a) and one with magnetic moments rotated by 90$^\circ$ in
the $c$-plane, is small and at the edge of the accuracy ($10~\mu$eV) of the
calculation in this particular case.

To model actual conditions in our experiments, we perform zero temperature
calculations of $E^{\mathrm{MCA}}$ for various choices of lattice constants
(see Tab.~\ref{tab-02}).  Adding the MCA to the dipole term, we can conclude
that (a) the out-of-plane MAE is typically between 0.2 and 0.3 meV per unit
cell (two formula units), favoring the moments within the $c$-plane, and (b)
the anisotropy within the $c$-plane is typically an order of magnitude smaller.
For calculations under changing $c/a$-ratio shown in Tab.~\ref{tab-02}, the MAE
within the $c$-plane is always smaller than the out-of-plane MAE (even for the
extreme choice of lattice constants with $c=0.689$~nm, see Tab.~\ref{tab-02},
the latter is greater than 0.1~meV per unit cell), the MAE within the $c$-plane
exhibits no clear trend upon unit cell deformation and it even changes sign. In
order to unambiguously determine anisotropies within the $c$-planes, it is
therefore advisable to resort to experiments.

\begin{table}
\begin{tabular}{l|cccccc}
$a$ [nm] 				& 0.408 & 0.411 & 0.414 & 0.417 &  0.408 & 0.408 \\ 
$c$ [nm] 				& 0.670 & 0.670 & 0.670 & 0.670 &  0.650 & 0.689 \\ \hline
$E_{[0001]}-E_{[11\bar{2}0]}$ 		& 0.20    & 0.24        & 0.23        & 0.22 & 0.28 &  0.12\\
$E_{[1\bar{1}00]}-E_{[11\bar{2}0]}$ 	& $-0.01$ & 0.03        & 0.01        & 0.04 & 0.05 & $-0.01$ 
\end{tabular}
\caption{The total MAE, $E^{\mathrm{dipole}}+E^{\mathrm{MCA}}$ in meV per unit
cell for different lattice parameters.  The N\'eel vector directions with
respect to the crystal are given as subscript of the energies, showing the
preferential magnetic moment orientation in the $c$-plane.}
\label{tab-02}
\end{table}

\section{Sample structure and magnetometry}

\subsection{Sample growth and structure}

\begin{figure}
\includegraphics[width=\linewidth]{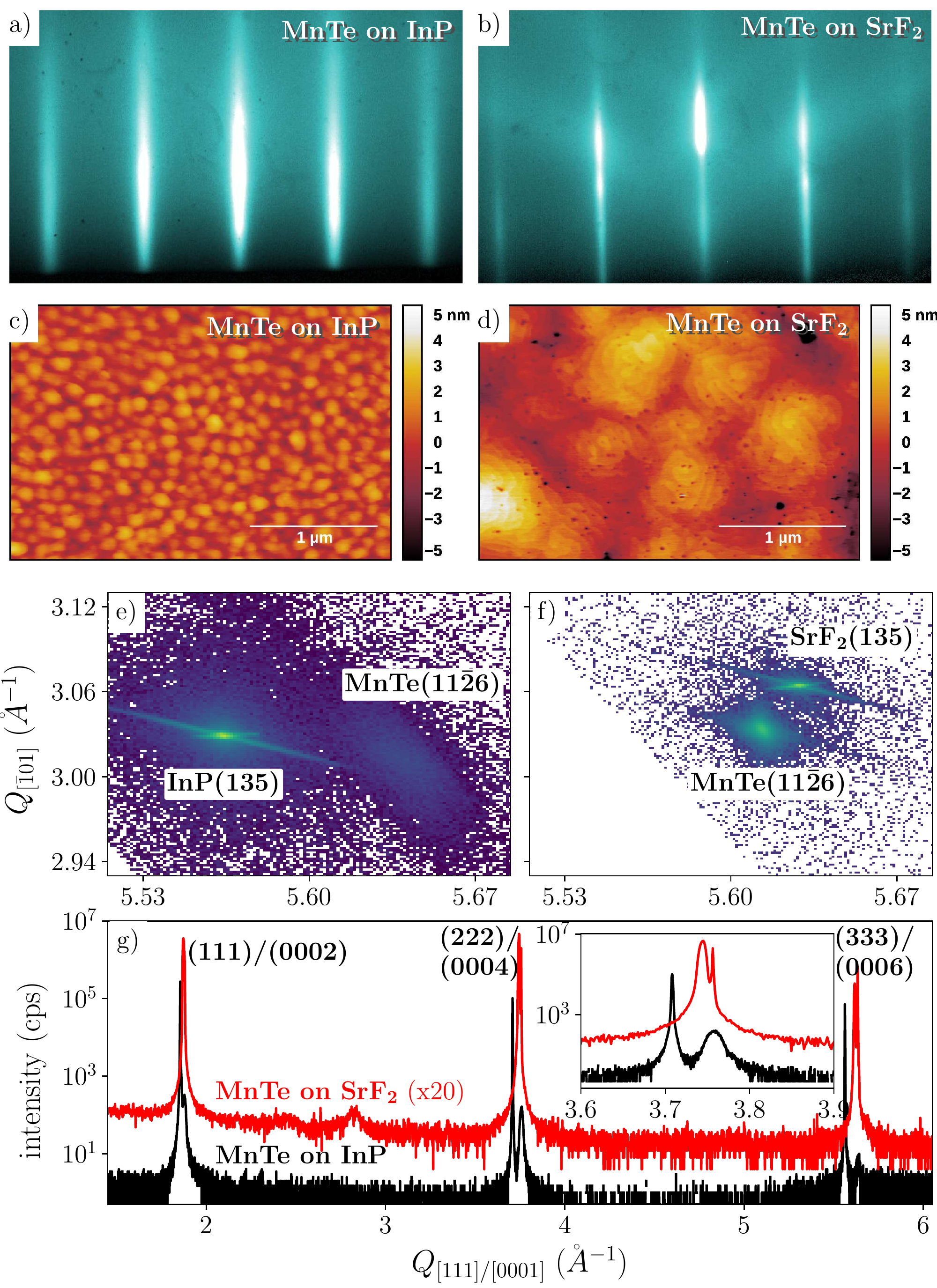}
  \caption{Reflection high-energy electron diffraction (RHEED) patterns of
  50~nm and 2500~nm thick epitaxial MnTe films on InP(111)A (a) and
  SrF$_2$(111) (b), respectively, recorded along the $[\bar110]$ zone axis.
  Atomic force microscope images for MnTe films on InP(111)A and SrF$_2$(111)
  are shown in panels (c,d).  X-ray diffraction reciprocal space maps around
  the (135) substrate Bragg reflection of the samples are shown in (c) and (d).
  Due to the epitaxial relationship the $(11\bar26)$ Bragg reflection of MnTe
  is seen close to the substrate peaks. Panel (e) shows the symmetric radial
  scan for MnTe on InP (black) and MnTe on SrF$_2$ (red). For clarity the trace
  of MnTe on SrF$_2$ was scaled by a factor of 20. The inset shows a zoom
  around the (222) substrate Bragg peaks and the broader (0004) Bragg peaks of
  the MnTe epilayer. Data in panels (c-e) are recorded at room temperature
  where due to different strain states the Bragg peak position of the films is
  slightly different for the two used substrate materials.}
\label{fig:rheed}\label{fig:xrd}\label{fig:2}
\end{figure}

Single crystalline hexagonal MnTe epilayers were grown by molecular beam
epitaxy on single crystalline SrF$_2$(111) and In-terminated InP(111)A
substrates using elemental Mn and Te sources. Both types of substrates have a
cubic structure (CaF$_2$ and zinc blende, respectively) with lattice parameters
of $a_0=0.57996$ nm for SrF$_2$ and 0.58687 nm for InP at room temperature.
However, the respective surface lattice constants of the (111) surfaces
($a_0/\sqrt{2}$) of 0.410~nm and 0.415~nm are very close to the hexagonal $a$
lattice constant of $\alpha$-MnTe ($a=0.414$~nm and $c=0.671$~nm
\cite{Greenwald1953}).  Thus, both types of substrates are very well suited for
MnTe growth with a lattice mismatch of less than 1\% in both cases which
resulted in single crystalline films with epitaxial interface between film and
substrate~\cite{Kriegner2016}. Indeed two-dimensional growth of $\alpha$-MnTe
is achieved at the used substrate temperatures in the range of 370-450$^\circ$C
as indicated by the streaked reflection high-energy electron diffraction
(RHEED) patterns observed during growth as shown in Fig.~\ref{fig:rheed}(a,b).
The resulting film morphology can be seen in the atomic force microscopy images
in Fig.~\ref{fig:rheed}(c,d). In both cases a root mean square roughness of
$\sim1$~nm is observed.  From X-ray diffraction measurements shown in
Fig.~\ref{fig:rheed}(e-g), we find that the MnTe layers grow with the $c$-axis
perpendicular to the (111) substrate surfaces with an epitaxial relationship of
$(0001)[1\bar100]_{\rm MnTe} || (111)[11\bar2]_{\rm SrF2/InP}$, which
corresponds to hexagon-on-hexagon like matching. Thus we refer to the $c$-axis
as out-of-plane direction and all perpendicular directions including
$\left<10\bar10\right>$ and $\left<11\bar20\right>$, i.e.  within  the
$c$-plane, are called in-plane directions. Note that we use the Miller indices
$hkl$ to denote cubic and Bravais indices $hkil$ with $i = -h-k$ to denote the
hexagonal lattice points/directions and that crystallographic directions refer
to the corresponding real space directions. In addition, no traces of secondary
MnTe phases are found in the wide range diffraction scans in
Fig.~\ref{fig:rheed}(g).  

From reciprocal space maps as shown in Fig.~\ref{fig:xrd}(e,f) the in-plane and
out-of-plane lattice constants of the epilayers $a$ and $c$ were determined.
For all MnTe films on SrF$_2$ (111) thicker than 50 nm, we generally find that
the in-plane lattice constant is very close the MnTe bulk value indicating
that the films are nearly fully relaxed.  On the contrary, the films on InP
(111) exhibit an in-plane lattice constant larger than bulk MnTe in spite of the
closer lattice matching.  This is explained by the different thermal expansion
coefficients of the film and the substrate. Above room temperature, the
thermal expansion coefficient of SrF$_2$ is $2.0\times10^{-5}$K$^{-1}$ (Ref.
\onlinecite{Kommichau1986}), which is only 20\% larger than the
value of $1.62\times10^{-5}$K$^{-1}$ (Ref. \onlinecite{Minikayev2015}) of MnTe,
for which reason the cooling of the sample from the growth temperature to
room temperature does not induce a significant thermal strain in the films due
to nearly the same thermal contraction. Conversely, the thermal coefficient of
InP of $0.5\times10^{-5}$K$^{-1}$ (Ref. \onlinecite{Glazov1977}) is about three
times smaller than that of MnTe and therefore, the cooling to room temperature
induces a significant tensile strain in the epilayers exceeding 0.5\%. Thus,
MnTe films on InP are subject to bi-axial tensile strain whereas
those on SrF$_2$ are
nearly fully relaxed.  For $\mu$m thick MnTe films of InP, the large thermal
expansion mismatch stress leads to the formation of microcracks in the films as
well as partial delamination.  For this reason, only thick films on SrF$_2$
were used for our neutron diffraction investigations. For the investigations of
the magnetic anisotropy of the films on InP the films thickness was restricted
to 50 nm and therefore magneto transport measurements instead of neutron
diffraction were used.

\begin{figure}
\includegraphics[width=\linewidth]{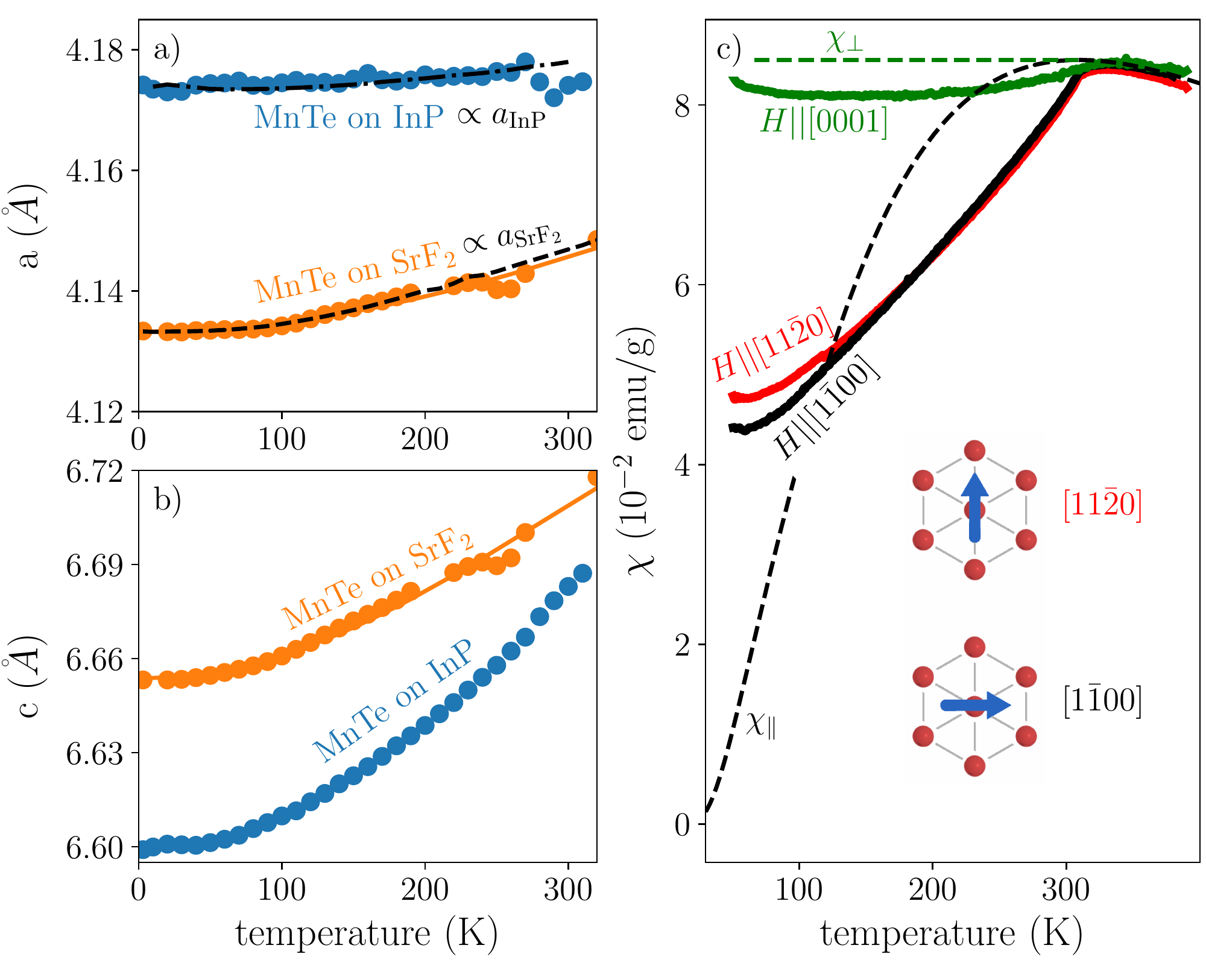}
  \caption{(a,b) Temperature dependent $a$ and $c$ lattice parameters of MnTe
  grown on two different substrates. The dashed (dash-dot) line in (a)
  represents the measured temperature dependence of the SrF$_2$ (InP) substrate
  lattice parameter scaled by $\sqrt{2}*1.01$ ($\sqrt{2}*1.007$). Note that the
  solid lines shown for the case of MnTe on SrF$_2$ are guides to the eye since
  measurements around room temperature are hampered by overlapping signals of
  the thin film and substrate. (c) Temperature dependent susceptibility of
  2.5~$\mu$m MnTe on SrF$_2$ measured for magnetic field applied in different
  directions. The diamagnetic contribution of the substrate was subtracted.
  Dashed lines show the mean-field 
  susceptibility of a collinear uniaxial antiferromagnet for the cases when the
  field is perpendicular ($\chi_\perp$/green) and parallel
  ($\chi_\parallel$/black) to the easy axis.  Insets indicate the directions of
  the magnetic field with respect to the crystal within the $c$-plane.}
\label{fig:xrdtemp}\label{fig:suscept}\label{fig:3}
\end{figure}

The different thermal expansion of the layers and substrates will also
modify the strain state of the MnTe films at low temperatures (neutron
diffraction and magnetotransport investigations are performed at liquid Helium
temperatures). Therefore, we measured in addition the temperature dependence of
the lattice parameters by X-ray diffraction as shown in
Fig.~\ref{fig:xrdtemp}(a,b).  When cooled from room-temperature, the in-plane
lattice constant $a$ of both films on SrF$_2$ and InP basically follows the
change of the scaled substrate lattice parameter which is also plotted in
Fig.~\ref{fig:xrdtemp}(a) by the dashed and dash-dotted lines. This
means that the in-plane
strain of the MnTe film on InP even increases, whereas only small changes
occur on SrF$_2$. Note that the scaling of the substrate surface lattice
parameters by around 1\% indicates the relaxation of the epitaxial films during
growth.  At liquid He temperatures the in-plane lattice constant of the MnTe
films differs by as much as 1.0\% for the different substrates.  This leads also
to a different evolution of the out-of-plane $c$ axis lattice constant of the
films on InP and SrF$_2$ as shown in Fig.~\ref{fig:xrdtemp}(b).  Our
theoretical calculations in Tab.~\ref{tab-02} indicate that while the
out-of-plane MAE remains dominant upon such variations of strain, the in-plane
MAE may change substantially, potentially even to the point that the direction
of the easy axis (within the basal plane) changes. 

\subsection{Magnetometry}

One possible way of determining the natural orientation of magnetic moments,
i.e.  the easy axis direction, is the measurement of the temperature-dependent
susceptibility $\chi$ shown in Fig.~\ref{fig:suscept}(c). Very early
on,\cite{Nagamiya1955} it has been recognized that while $\chi_\parallel(T)$
(magnetic field applied parallel to magnetic moments) for a uniaxial
antiferromagnet drops to zero as $T\to 0$, magnetic field applied in (any)
perpendicular direction gives a constant $\chi_\perp(T)=\chi_0$
for $T<T_{\rm N}$. Explicit form of $\chi_0$ as well as 
$\chi_\parallel(T)$ based on Weiss theory can be found in
Ref.~\onlinecite{Baltz2017}.
We show this mean-field theory result for $S=5/2$ and scaled to the experimental
data in Fig.~\ref{fig:suscept}(c) as dashed lines.
Experimental data for $H||[0001]$
therefore confirm that magnetic moments lie in the basal plane. On the other
hand, since neither of the other two curves for $H||[11\bar20]$ and
$H||[1\bar100]$ approaches zero for low temperatures, we conclude that there
is not one single easy axis (or in other words, the sample is not uniaxial and
therefore not in a single domain state). The small difference between these two
curves suggests that the anisotropy within the $c$-planes is small. 

\section{Neutron diffraction investigations}

\begin{figure}
\includegraphics[width=\linewidth]{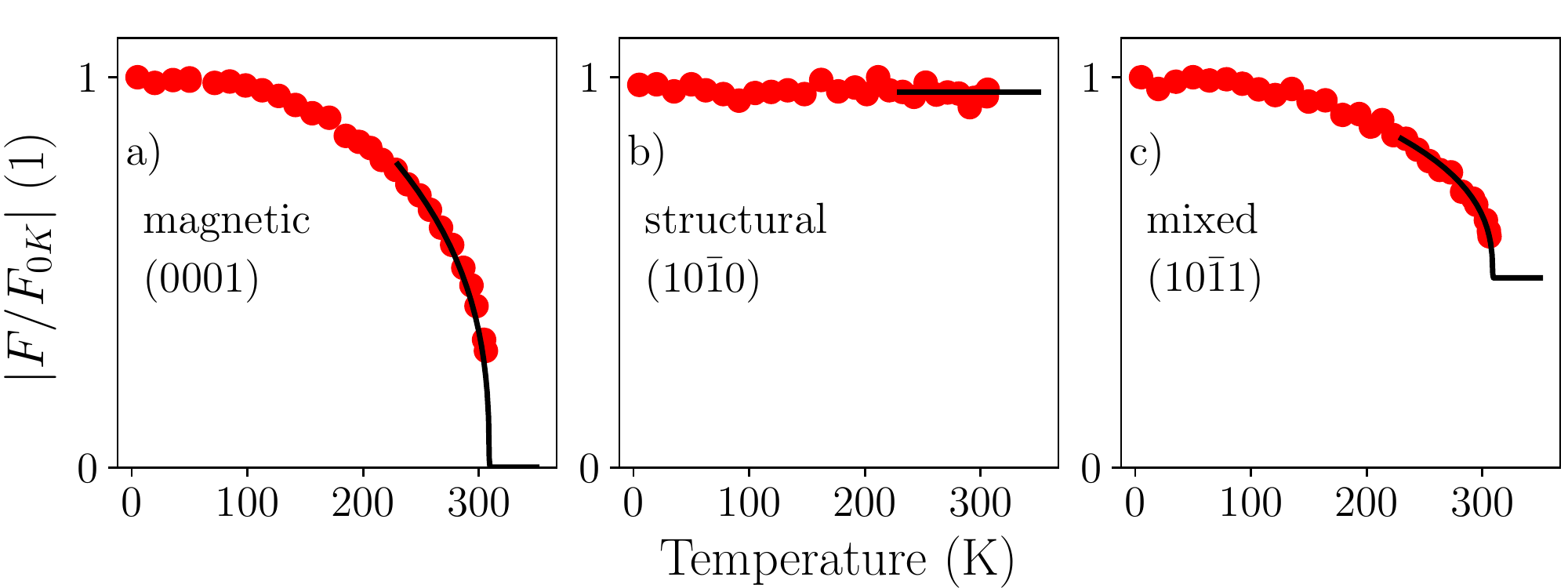} 
\caption{Neutron diffraction structure factors measured versus temperature for
the (a) magnetic (0001), (b) structural $(10\bar10)$, and (c) mixed structural
and magnetic Bragg peak $(10\bar11)$. Black lines show the behavior close to
the N\'eel temperature described by the equation $A_{\rm mag} \left(T_{\rm N}
- T \right)^{c} + A_{\rm struct}$, where $A_i$ denotes the amplitude of
magnetic and structural contribution and $T$ the measurement
temperature. The N\'eel temperature $T_{\rm N}=309$~K and the
critical exponent $c=0.37$ corresponds to the Heisenberg model.\cite{Chen1993}}
\label{fig:neutrontemp}\label{fig:4}
\end{figure}

Experiments at the CEA-CRG thermal neutron diffractometer D23 at Institut
Laue-Langevin in Grenoble, France allowed us to determine the easy axis in MnTe
layers grown on SrF$_2$. A monochromatic beam of neutrons with a wavelength of
0.127~nm was generated by a Cu (200) monochromator. The sample was mounted in a
rotatable cryomagnet with temperature range of 5 to 305~K and magnetic fields
up to 6~T along the sample rotation axis. The diffraction geometry with two
orthogonal rotation axes of the detector allowed to access several MnTe Bragg
peaks sufficiently separated from those of the
substrate. In Fig.~\ref{fig:neutrontemp}, we show the intensity of selected
diffraction peaks as a function of temperature. Since non-polarized neutrons
were used, the magnetic diffraction intensity depends solely on the relative
orientation of the magnetic moments and the momentum transfer, and is at
maximum when the magnetic moment is perpendicular to the momentum transfer.
The shown variation of the (0001) diffraction peak
(Fig.~\ref{fig:neutrontemp}(a)), which is structurally forbidden in the
paramagnetic phase, indicates that the magnetic moment within the $c$-plane
has a significant value. In contrast to that a peak with momentum transfer
within the $c$-plane (see ($10\bar10$) in Fig.~\ref{fig:neutrontemp}(b)) shows
no magnetic contribution and therefore its intensity is virtually independent
of temperature. The variation of the structure factors close to N\'eel
temperature can be described by the critical behavior of the Heisenberg
model with exponent $c=0.37$\cite{Chen1993}
and is shown as solid line in
Fig.~\ref{fig:neutrontemp}. The ratio of intensities of the purely structural
and magnetic Bragg peaks can therefore be used to determine the magnetic moment
of the Mn atoms. By comparison with simulations using the FullProf Suite
\cite{Rodriguez1993} we find a magnetic moment between 4.7 and 5 $\mu_B$ at
low temperature, which is in agreement with earlier studies\cite{Kunitomi1964,
Szuszkiewicz2005}.

\begin{figure}
\includegraphics[width=\linewidth]{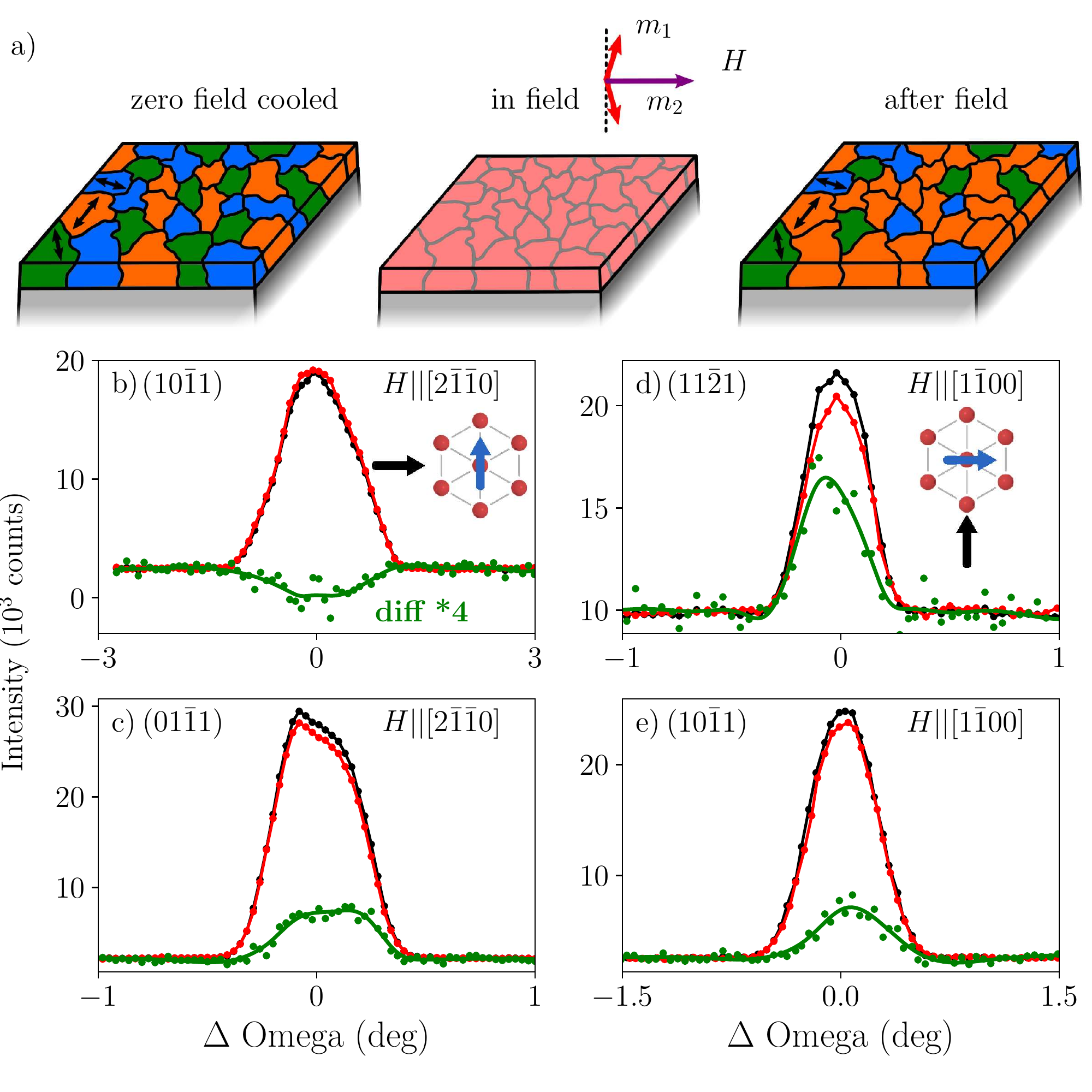}
  \caption{(a) Effect of a strong magnetic field on the
  domain population in MnTe thin films. After zero field cooling the
  population of the three distinct magnetic domains is equal. When a strong
  field, {\it i.e.} above the spin-flop field, is applied the canted magnetic
  moments (see inset) of all domains align almost perpendicular to the magnetic
  field. After releasing the field a higher population of the domain with the
  easy axis close to perpendicular to the field direction remains (orange
  domains in the sketch). (b-e) Neutron diffraction curves recorded at
  a temperature of 5~K after zero field cooling (black) and after application
  and removal of a magnetic field of 6~T (red). The difference between the
  black and red curves is shown multiplied by a factor of 4 in green. The
  effect of the magnetic field is shown for two field directions differing by
  90~degrees. Insets in (b), and (d) show to orientation of the magnetic field
  (blue arrow) within the $c$-plane and indicate the direction of the primary
  neutron beam (black arrow).}
\label{fig:neutronint}\label{fig:5}
\end{figure}

Such intensity ratios, however, cannot be directly used to determine the
in-plane orientation of the magnetic moments. When the sample is cooled in zero
magnetic field magnetic domains \emph{equally} populate the various equivalent
crystallographic directions\cite{Komatsubara1963, Kriegner2016} ({\it cf.}
Fig.~\ref{fig:5}(a)) and the sample appears to be isotropic in the $c$-plane.
To break this symmetry one can apply a strong in-plane magnetic field above the
spin-flop transition, to enforce domain repopulation\cite{Kriegner2016}.
Such a field forces the moments in an orientation nearly perpendicular to the
applied magnetic field and therefore the magnetic diffraction intensities also
do not contain the desired information about the in-plane easy axis.  However,
when the strong applied field is removed the domains with easy axis direction
closest to perpendicular to the field direction are preferentially populated.
For the case when an in-plane field is applied perpendicular to one of the easy
axis this means that domains with this N\'eel vector orientation will be
preferred over the two other domains with N\'eel vector orientation tilted by
30$^\circ$ with respect to the field direction. From the difference of the
domain repopulations for various magnetic field directions one can determine
the easy axes directions. Below we show that a magnetic field of 6~T is
sufficient to repopulate the domains since it triggers the spin-flop
transition. Neutron diffraction measurements before (black) and after (red) the
application of a magnetic field for various Bragg peaks and two field
directions are shown in Fig.~\ref{fig:5}(b-e). As magnetic field directions we
use the high symmetry directions within the $c$-plane, namely the $[1\bar100]$
and $[2\bar1\bar10]$ directions.  Note that the measurements before and after
application of a magnetic field were performed on the very same sample, which
first was mounted so as to have the field direction along $[2\bar1\bar10]$
and removed after the measurement of Fig.~\ref{fig:5}(b,c), heated above N\'eel
temperature and remounted to have the field along $[1\bar100]$ to measure
Fig.~\ref{fig:5}(d,e). The difference between the measurements before and after
the field, i.e. the signal corresponding to the remnant domain repopulation,
are shown as the green curves. It is this difference which will be further
quantified and analyzed in Tab.~\ref{tab-01}.

Structure factors $F$ were extracted from the measurements using the software
COLL5\cite{Lehmann1974}, which considers geometrical effects from the
measurement setup, resulting in different full width at half maximum values
for different Bragg reflections shown in Fig.~\ref{fig:5}.  The relative
difference of structure factors before and after the application of magnetic
field $H$: $\Delta \left|F_{\rm EXP}\right| = \left(\left|F_{\rm
EXP}^\text{after H}\right| - \left|F_{\rm EXP}^\text{before
H}\right|\right)/\left|F_{\rm EXP}^\text{before H}\right|$ is listed in
Tab.~\ref{tab-01} for selected Bragg peaks and $[2\bar1\bar10]$ and
$[1\bar100]$ field direction.  In order to derive the easy axis direction we
further modeled the structure factors with FullProf for two different easy axes
directions. As potential easy axis directions we consider two high symmetry
directions: the $\left<1\bar100\right>$ direction indicated in
Fig.~\ref{fig:1}(a) and the direction perpendicular to it in-plane, i.e.
$\left<2\bar1\bar10\right>$. To derive the simulated change of the structure
factor $\Delta \left|F_{\rm SIM}\right|$ we additionally model the efficiency
of the domain repopulation after the application of a 6~T magnetic field. 

As mentioned above the magnetic field leads to higher population of the
domain(s) with easy axis closer to the field normal. Since the efficiency of
this process is unknown we considered it a free parameter in our model. The
domain population is described by three occupation numbers, which add up to
unity. Each occupation number corresponds to the occupation of a domain with
N\'eel vector orientation along one of the three crystallographically
equivalent axis within the $c$-plane. Taking into account the field directions
and considered easy axis directions this means that we either equally favor or
disfavor two sets of domains. This means that one parameter is sufficient to
describe the domain repopulation in either case. Since the two different field
directions with respect to the easy axes directions likely result in different
domain repopulation efficiencies, this means we have two free parameters in the
model.  Within this model the observed changes of the structure factors $\Delta
\left|F_{\rm EXP} \right|$ in Tab.~\ref{tab-01} can only be consistently
explained when we consider the easy axes to be along the
$\left<1\bar100\right>$ directions (cf. $\Delta \left|F_{\rm SIM}^{\left< 1
\bar1 0 0 \right>}\right|$ in Tab.~\ref{tab-01}).  The two free parameters
describing the domain population thereby result in populations of
$\sim40:30:30$\% and $\sim39:39:22$\% for the three distinct easy axes
directions after the application of the field perpendicular and parallel to one
easy axis. In Fig.~\ref{fig:5}(a) the change of the domain population by the
application of a field perpendicular to an easy axis, which leads to the
increase of one population, and corresponding decrease of the population of the
two other domains is qualitatively sketched. In agreement to
Ref.~\onlinecite{Kriegner2016} a single domain state is unachievable at least
after removal of the magnetic field.  The determined easy axes are consistent
with the susceptibility data measured by SQUID (cf.  Fig.~\ref{fig:3}(c)),
which found the lowest susceptibility at low temperature when the field is
aligned along the $\left[1\bar100\right]$ direction, or any other equivalent
direction.

\begin{table}
\caption{\label{tab:fabs} Relative change of the absolute value of the
structure factor after the application of a 6~T field in the specified
direction. The respective experimental data are shown in
Fig.~\ref{fig:neutronint}. The simulated change for easy axes along $\left< 1
\bar1 0 0 \right>$ and $\left< 2 \bar1 \bar1 0 \right>$ is listed and the
former simulation (highlighted) agrees within error bars with
experimental data.}
\begin{tabular}{cccccc}
peak            & $H$ direction                         & $\Delta \left|F_{\rm EXP} \right|$ (\%) & $\Delta \left|F_{\rm SIM}^{\mathbf{\left< 1 \bar1 0 0 \right>}} \right|$ & $\Delta \left|F_{\rm SIM}^{\left< 2 \bar1 \bar1 0 \right>} \right|$ \\ \hline
$(1 0 \bar1 1)$ & $\left[ 2 \bar1 \bar1 0 \right]$      & $1.27 \pm 0.16$ & {\bf 1.30\%} & 1.30\% \\
$(0 1 \bar1 1)$ & $\left[ 2 \bar1 \bar1 0 \right]$      & $-2.60 \pm 0.08$ & {\bf -2.60} & -2.60 \\
$(1 1 \bar2 1)$ & $\left[ 1 \bar1 0 0 \right]$          & $-7.23 \pm 0.30$ & {\bf -7.24} & -7.21 \\
$(1 0 \bar1 1)$ & $\left[ 1 \bar1 0 0 \right]$          & $-2.11 \pm 0.08$ & {\bf -2.03} & -1.22 
\end{tabular}
\label{tab-01}
\end{table}

\section{Magnetotransport}

\begin{figure}
\includegraphics[width=\linewidth]{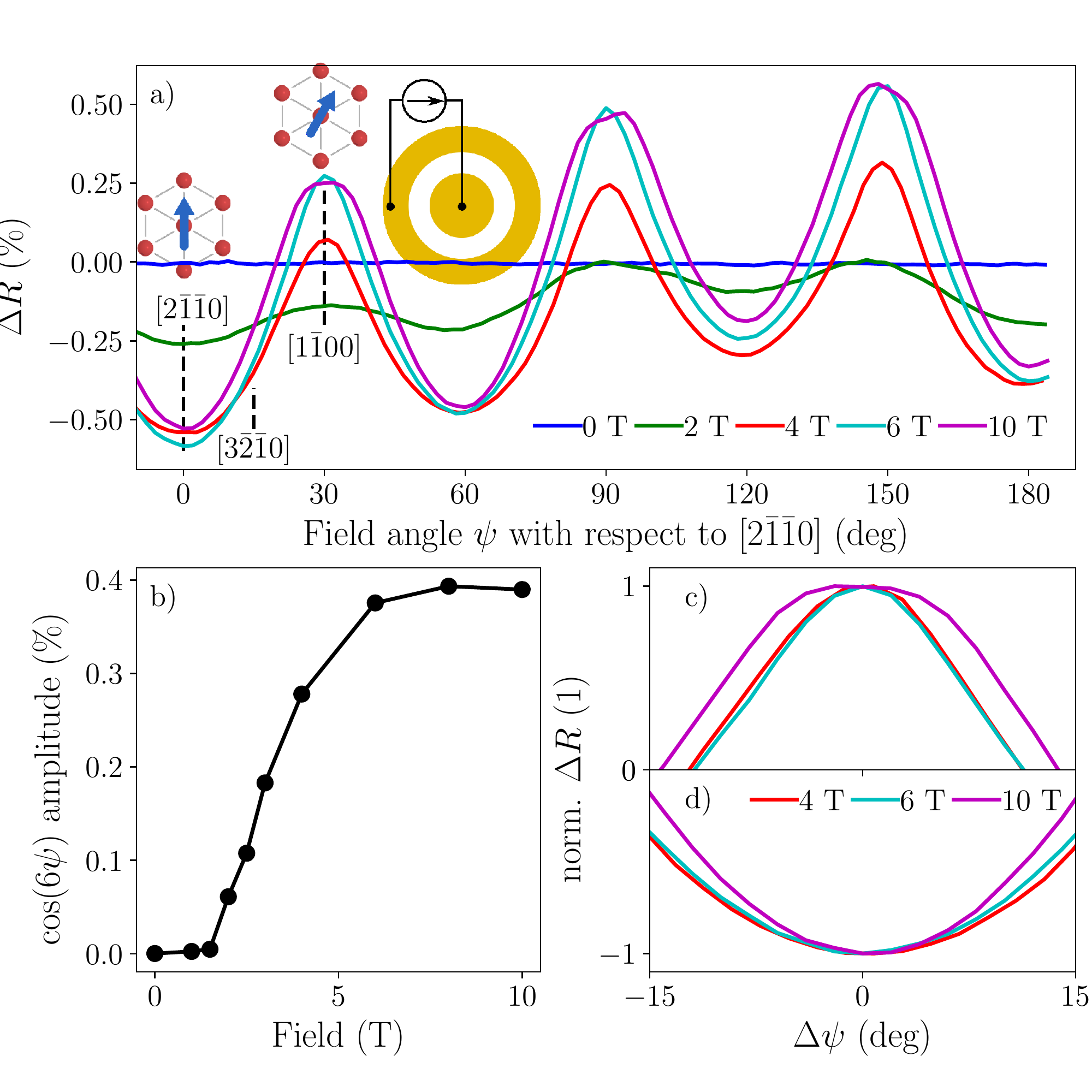}
  \caption{(a) Longitudinal resistance traces during magnetic field rotations in
  50 nm MnTe on InP(111) for different field strengths. An inset shows the
  Corbino disk measurement geometry, and three magnetic field directions are
  marked by their crystallographic directions. (b) Variation of the amplitude
  of the dominant $\cos(6\varphi)$ contribution to the resistance change. The
  amplitude was determined using a Fourier decomposition of the measured
  resistance change. (c,d) Zooms to the minimum and maximum resistance values
  during the field rotations. Maxima in (c) are narrower than minima in (d).}
\label{fig:transport}\label{fig:6}
\end{figure}

Since thick enough films for neutron diffraction cannot be obtained for MnTe
on InP(111) we employed an alternative approach to determine the easy axis
directions in this case. Using the crystalline contribution\cite{DeRanieri2008}
to the anisotropic magnetoresistance (AMR) the easy axis can also be
determined. Radial flow of electrical current in Corbino disks suppresses the
non-crystalline components\cite{Rushforth2007} and the remaining crystalline
contribution $\propto \cos(6\phi)$ due to the hexagonal symmetry of the
material serves as a straightforward detector of the N\'eel vector direction.
Corbino contacts, sketched in the inset of Fig.~\ref{fig:transport}(a),
were fabricated on MnTe thin layers (50~nm thick) grown on InP by
depositing gold contact rings using a lithographic lift-off process.

During an in-plane rotation of applied magnetic field, also when its 
strength is above the spin-flop threshold, the anisotropy
makes the N\'eel vector lag behind
the direction perpendicular to the field when the former is located near an
easy axis. Consequently, deviations from the $\cos(6\phi)$ form can be
observed in Fig.~\ref{fig:transport}. This means that, as soon as the
N\'eel order can be influenced by external magnetic field,
the easy axis can be determined from such transport measurements.
In Fig.~\ref{fig:transport}(a) we show the
field direction dependence of the longitudinal resistance for magnetic fields
up to 10~T. While at low fields almost no effect of a field rotation is
observed, a dominantly six-fold signal arises in stronger fields.
Figure~\ref{fig:transport}(b) shows the variation of the $\cos(6\psi)$
contribution to the AMR signal for different field strengths.  The mentioned
contribution shows a clear onset just below 2~T and saturates for fields above
6~T, indicating that all moments rotate slightly canted aligned almost
perpendicular to the stronger fields. We note that the six-fold variation of
the resistance shows clear differences between the maxima and minima of the
resistance variation. As it is visible in
Fig.~\ref{fig:transport}(c,d) the minima
always appear wider than the maxima.  This indicates that magnetic moments
are pushed towards the position of the minima in the resistance by the in-plane
anisotropy. Considering that the magnetic field is nearly perpendicular to the
moments we infer that the easy axes are oriented along
$\left<1\bar100\right>$. Note that the difference between minima and maxima
is decreasing in stronger fields as the MA is becoming smaller
relative to the external magnetic field.

\section{Spin-flop field measurements}

\begin{figure}
\includegraphics[width=\linewidth]{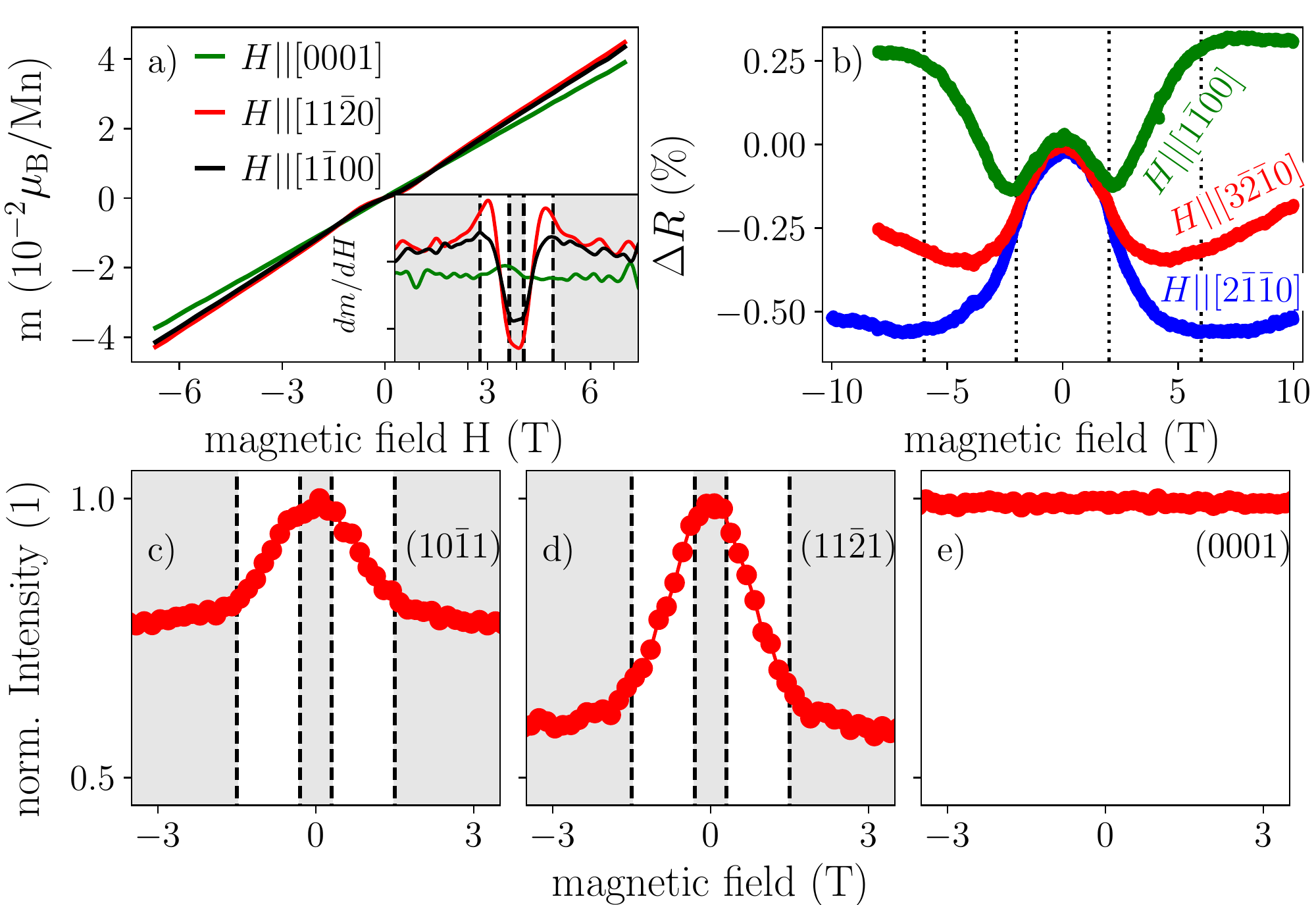}
  \caption{Magnetic field sweeps performed using different methods. (a)
  Magnetic moment per Mn atom of MnTe on SrF$_2$ determined by SQUID for
  magnetic fields applied in various directions. The diamagnetic contribution
  of the substrate was subtracted. An inset shows the derivative of the
  magnetic moment by the magnetic field. For the in-plane measurements distinct
  regions are detected which are distinguished by the slope of $m(T)$.
  (b) Change of the longitudinal
  resistance in MnTe on InP(111) measured in Corbino disk geometry vs.
  magnetic field applied along the $[2\bar1\bar10]$, $[3\bar2\bar10]$, and
  $[1\bar100]$ directions.  (c-e) Normalized neutron diffraction intensity of
  the $(10\bar11)$, $(11\bar21)$, and (0001) Bragg peak of MnTe grown on
  SrF$_2$ during a magnetic field sweep with field along the $[1\bar100]$
  direction.}
\label{fig:7}
\end{figure}

In Fig.~\ref{fig:7} we plot magnetic field dependent measurements, revealing
the spin-flop transition detected by various methods using both considered
substrate materials.  Figure~\ref{fig:7}(a) shows the magnetization per Mn atom
measured by a SQUID magnetometer when strong magnetic fields are applied. As
expected for an antiferromagnet, the magnetization of the sample is mostly
compensated and only a fraction of Bohr magneton $\mu_B$ is detected even
above the spin-flop transition.  When the field is applied in the out-of-plane
$c$-direction, a featureless linear trace is observed, while for in-plane field
small changes of the slope appear (best visible in the inset), indicating the
spin-flop transition.  Since our system comprises multiple domains and three
in-plane easy axis directions the traces deviate from the more common spin-flop
signals in uniaxial antiferromagnets~\cite{Jacobs1961, Oh2014}. However, the
characteristic features with smaller slope below the spin-flop field and a
higher slope above the spin-flop field are clearly visible in our data.  Note
that as expected for an antiferromagnetic material, the slope of the traces at
high fields when extrapolated to zero crosses through zero, which excludes any
ferromagnetic contribution. The net magnetization of $\sim 0.04\mu_{\rm B}$/Mn
at the highest field of 6~T corresponds to a canting angle smaller than
1$^\circ$.

Field dependent neutron diffraction intensities shown in panels
Fig.~\ref{fig:7}(c-e) confirm the spin-flop field as observed by the SQUID
magnetometer.  Similar to the SQUID measurements, different regions in
Fig.~\ref{fig:7}(c,d) can be identified (indicated by gray background color).
At small fields (below 0.5~T) and above $\sim1.5$~T the intensities are rather
constant while up to 40\% changes are observed between 0.5 and 1.5~T. This
shows that a certain field needs to be overcome to start the reorientation of
the moments.  Once the reorientation is complete no changes occur in the
neutron diffraction intensities since in contrast to SQUID neutrons are not
sensitive enough to detect the small magnetic moment induced by the canting of
the two magnetic sublattices.  It is important to note that the magnetic
diffraction peak (0001) is unaffected because the magnetic moments remain in
the basal plane and therefore are always perpendicular to the momentum
transfer. This again confirms that the [0001] direction is the hard axis of the
system in agreement with our theoretic predictions and magnetic susceptibility
measurements in Fig.~\ref{fig:suscept}.

Magnetic field sweeps in transport measurements shown in Fig.~\ref{fig:7}(b)
also show significant changes associated with the spin-flop.  Instead
of the reorientation of moments between $\sim0.5$ and 1.5~T, as seen by neutron
diffraction and SQUID, the onset of AMR in these measurements is located
between $\sim2$ and 6~T. This large change implies that MA in both
samples are different since the spin-flop field is proportional to square root
of the MAE.\cite{Bernstein1971} Note that the sample used in these transport studies
was grown on InP(111) which causes different strain. Although the easy axis
directions determined from neutron diffraction for films grown on SrF$_2$ are
found to be the same as the ones determined from AMR for films on InP, the
strength of the in-plane anisotropy is different. On InP the tensile strain in
MnTe at low temperature (mostly due to thermal expansion coefficient
mismatch) causes a bigger magnetic anisotropy, resulting in a higher spin-flop
field for those samples. 

\section{Conclusion}

Our neutron diffraction, magnetometry, and magnetotransport measurements in
combination with DFT+U calculations confirm that antiferromagnetic
NiAs-type MnTe
thin layers are magnetically an easy plane material. Within the hexagonal basal
plane, the magnetic anisotropy is considerably smaller than the out-of-plane
anisotropy and moreover, can be engineered by choosing suitable substrate and
working temperature. For MnTe films on InP and SrF$_2$ at low temperatures, the
easy axis is $[1\bar{1}00]$ (or any of the other two crystallographically
equivalent directions). The strain induced by the thermal expansion coefficient
mismatch on InP causes tensile strain within the $c$-plane and results in
significantly higher spin-flop fields. Onsets of the spin-flop transition
change from $\sim0.5$~T for films grown on SrF$_2$ to $\sim2$~T for films grown
on InP. The moderate spin-flop field allows to repopulate magnetic domains even
in the antiferromagnetic state, which was exploited to determine the easy axis
direction from neutron diffraction. The small in-plane anisotropy opens up the
possibility to vary the resistance of the material almost continuously due to
the AMR effect\cite{Kriegner2016}.  This together with its simple collinear
magnetic structure makes MnTe a favorable model system to test
antiferromagnetic spintronics phenomena. 

\begin{acknowledgments}
We acknowledge support from the Austrian Science fund (J-3523-N27), the Grant
Agency of the Czech Republic (grant no.  14-37427G), ERDF (project
''Nanomaterials centre for advanced applications'',
CZ.02.1.01/0.0/0.0/15\_003/0000485), the Ministry of Education of the Czech
Republic Grants No. LM2015087 and No. LNSM-LNSpin, the EU FET Open RIA Grant
No. 766566, and the ERC Synergy Grant No. 610115.
\end{acknowledgments}

%

\end{document}